# New eruptive variable in the massive star-forming region associated with IRAS 18507+0121 source

E.H. Nikoghosyan, N.M. Azatyan, and K.G. Khachatryan

Byurakan Astrophysical Observatory,
 Armenia, e-mail: lena@bao.sci.am

**ABSTRACT**

*Aims.* We report the discovery of a strong outburst of the embedded young stellar object (YSO), namely UKIDSS-J185318.36+012454.5, located in the star-forming region associated with IRAS 18507+0121 source and GAL 034.4+00.23 HII region.
*Methods.* Using the archival photometric data and images we determined the amplitude and the epoch of the outburst, as well as the evolution stage and the basic parameters of the object.
*Results.* According to the near and mid-infrared colors and spectral energy distribution, we classify the object as an intermediate mass young stellar object (YSO) with Class 0/I evolution stage. The outburst occurred in the period of 2004-2006. The amplitude of brightness is as least $K_s = 5.0$ mag. The summation of the photometric and spectral data does not allow to classify UKIDSS-J185318.36+012454.5 as FUor or EXor.  We can consider it as an eruptive variable with mixed characteristics or MNor type object.

## 1. Introduction

Eruptions of pre-main sequence (PMS) stellar objects are rare events, thus a new outburst is always noteworthy. This topic always has been in the spotlight of observers and theorists. Nevertheless, the number of verified eruptive variable PMS stars remains too small. However, due to the significant developments of infrared observations, that allows penetrating in embedded and distant star-forming regions, there has been some progress in this regard (Scholz et al., 2013; Contreras Peña et al., 2014, 2016). Most sun-like eruptive variable YSOs have been classically divided into two sub-classes, namely, FUors and EXors. FUors eruptive variables show a larger increase of luminosity (up to 6 magnitudes) and slow decay over (> 10 yrs) (Hartmann & Kenyon, 1996). This class of objects generally displays strong CO and $H_2O$ absorption and mostly lack of emission lines in their infrared spectra. EXors type variables have recurrent short-lived  (as rule < 1yr) outbursts (Herbig, 2008). In contrast with FUors a strong Bry and CO emission is dominated during the outburst in an infrared spectrum of EXors. These optically defined classification does not include younger embedded protostars that have higher accretion rates and a mixture of characteristics for which a relatively new label "MNOrs" is proposed (Contreras Peña et al., 2014).

In this work we report the discovery of a new embedded eruptive variable YSO UKIDSS-J185318.36+012454.5 (DR6 release of UKIDSS GPS survey). The stellar object is located in the massive star-forming region associated with IRAS 18507+0121 source. In the vicinity of IRAS source an ultra-compact (UC) H II region (G034.4+00.23) embedded in the filamentary Dark

Molecular Cloud (MSXDC G034.43_00.24) was discovered (Miralles et al., 1994). UKIDSS-J185318.36+012454.5 is also identified in the RMS archive as G034.4035+00.2282A (Lumsden et al., 2013). Continual dust emission, traced by millimeter and submillimeter images, revealed in this star-forming region the compact clumps, which named MM1-MM4 (Rathborne et al., 2005). One of them, MM2, coincides with IRAS 18507+0121 star-forming region. The G034.43_00.24 MM2 clump traced also by $H^{13}CO^+$, SiO, 6-cm emission (Harju et al., 1998; Shepherd et al, 2004; Shepherd et al, 2007). The different manifestations of star-forming activity are detected in the region. The IRAS 18507 region is associated with variable $H_2O$ (Miralles et al, 1994), $CH_3OH$ (Szymczak et al., 2000), OH (Edris et al., 2007) and $NH_3$ (Molinari et al., 1996) maser emission. Three massive molecular outflows are centered on or near G34.4+0.23 region (Shepherd et al., 2007). The outflow elongated from north to south direction are detected also in mid-infrared wavelengths (Shepherd et al., 2007; Cyganowski et al., 2008).

The study of the stellar population identified in this star-forming region some massive protostars with an age of about $10^5$ years, as well as a low-mass stellar population with age $\sim 10^6$ years (Shepherd et al., 2007). The authors suggest that the stars in this region may have formed in two stages: first, lower mass stars were formed and then more massive stars began to form. UKIDSS-J185318.36 + 012454.5 is located at a distance $\sim 3''$ to the NW direction from the luminous of them (IR source #54 from Shepherd et al. 2004). In the immediate vicinity of the IRAS 18507+0121 source, the compact group of YSOs (including UKIDSS-J185318.36 + 012454.5) was revealed (Morales et al., 2013; Azatyan et al., 2016).

The distance estimations of G34.4+0.23 region are very contradictory. VLBI parallax measurements of $H_2O$ maser sources within an Infrared Dark Cloud MSXDC G034.43+00.24 determined a distance of 1.56 kpc (Kurayama et al., 2001). The estimations of a kinematic distance considerably exceed the previous value and vary from 3.7 kpc (Simon et al., 2006) to 3.9 kpc (Molinari et al. 1996). Although a parallax measurement would seem to be more reliable distance determination, the authors of Foster et al. (2012) suggested that the parallax determinations to the same sources are incorrect because of the low declination of this target.

In this paper we present the evidence of the outburst, which occurred in YSO UKIDSS-J185318.36 + 012454.5 or G034.4035+00.2282A. We have defined the amplitude and the epoch of the outburst, as well as an evolution stage and the basic parameters of the stellar object.

## 2. Data of observations

For identification and study of the eruptive variable UKIDSS-J185318.36 + 012454.5, we acquired the archival infrared and submillimeter data and images. For NIR bands the photometric data from DR6 of UKIRT Galactic Plane Survey (UKIDSS GPS, Lucas et al., (2008)) and Cooper et al. (2013), as well the K images from the ESO archive (SOFI infrared imager/spectrometer (Moorwood et al., 1998)), DR9 release of UKIDSS GPS and Varricatt et al. (2010) were used.

For mid-infrared bands, we acquired IRAC data (Fazio et al., 2004) from the Spitzer Science Archive, which include the photometric parameters obtained from GLIMPSE I (Benjamin et al. 2003; Churchwell et al., 2009) and DEEP GLIMPSE (Whitney et al., 2011) catalogues. In addition, for photometric measurements we used [3.6] and [4.5] bands images obtained from programs with AOKEYs 42718976, 10291200 and 42728960.

Photometric measurements were done according to the standard procedure using the aperture photometry package APPHOT in IRAF. The zero-point fluxes of [3.6] and [4.5] bands (Vega-standard magnitudes for 1 $DN\ s^1$) were adopted from Reach et al. (2005).

We used also the data from 2MASS (Skrutskie et al., 2006), DENIS (Epchtein et al., 1999), AKARI (Ishihara et al., 2010) and Hi-GAL (Molinary et al., 2016) surveys, as well as from RMS archive (Lumsen et al., 2013).

## 3. Results
### *3.1. Luminosity variability*

The photometric data and images of UKIDSS-J185318.36+012454.5 in near-infrared bands taken in different epochs are presented in Tables 1 and Fig. 1 and respectively. They testify, that the stellar object was not identified in 2MASS and DENIS surveys. Therefore, its brightness in 1999 and 2000 years was fainter than Ks = 18 magnitude. Additionally, it is necessary to note, that on K image in Shepherd et al. (2004) obtained in 1999 the object also was invisible. On the $K_s$ images obtained with higher sensitivity on United Kingdom Infrared Telescope in 2003, the stellar object is barely noticeable as the nebulous knot, the surface brightness of which does not exceed 18.5 mag. Then, during 2003-2006 the luminosity of the object strongly increased, at least on 5 mag. in $K_s$ band and reach ~13.75 mag. The object's brightness remained practically unchanged until 2011.

**Table 1.** NIR photometric data

| Source | Date of obs. | H [mag] | $K_s$ [mag] |
|---|---|---|---|
| 2MASS | 1999-08-10 | – | – |
| DENIS | 2000-07-03 | – | – |
| Varricatt et al. (2010) | 2003-05-27 | – | $K_s \geq 18.5$ |
| DR6 UKIDSS GPS | 2006-06-01 | 17.07 ± 0.04 | 13.76 ± 0.01 |
| Cooper et al. (2013) | 2007-06-02 | 17.20 ± 0.30 | 13.76 ± 0.08 |
| SOFI (ID 083.C-0846) | 2010-06-05 | – | 13.68 ± 0.10 |
| DR9 UKIDSS GPS | 2011-08-11 | – | 13.75 ± 0.03 |

In order to make sure that such significant increase in the brightness of this embedded stellar object is not conditioned by the variability of the extinction along the line of sight, we determined the Ks magnitudes of stars, located in the immediate vicinity from UKIDSS-J185318.36 + 012454.5. They are marked on Fig. 1. All objects identified in Shepherd et al. (2004) and UKIDSS GPS survey. The near-infrared photometric data of the neighboring stars presented in Table 2. Besides magnitudes from UKIDSS GPS database, we determined their Ks magnitudes on images, which are presented in Varricatt et al. (2010). Between these epochs, the significant brightness increasing of UKIDSS-J185318.36+012454.5 was detected. As seen from data in Table 2, the brightness variability of nearby stars is not significant and does not exceed ~0.5 mag. It can be seen as an argument in favor of the fact that such a significant brightness increasing of UKIDSS-J185318.36+012454.5 did not occur due to the extinction variability of the surrounding gas-dust matter. Moreover, ΔK ≈ 5 mag. requires ΔAv ≈ 50 mag., that is unlikely. Hence, we can consider UKIDSS-J185318.36+012454.5 as an eruptive variable.

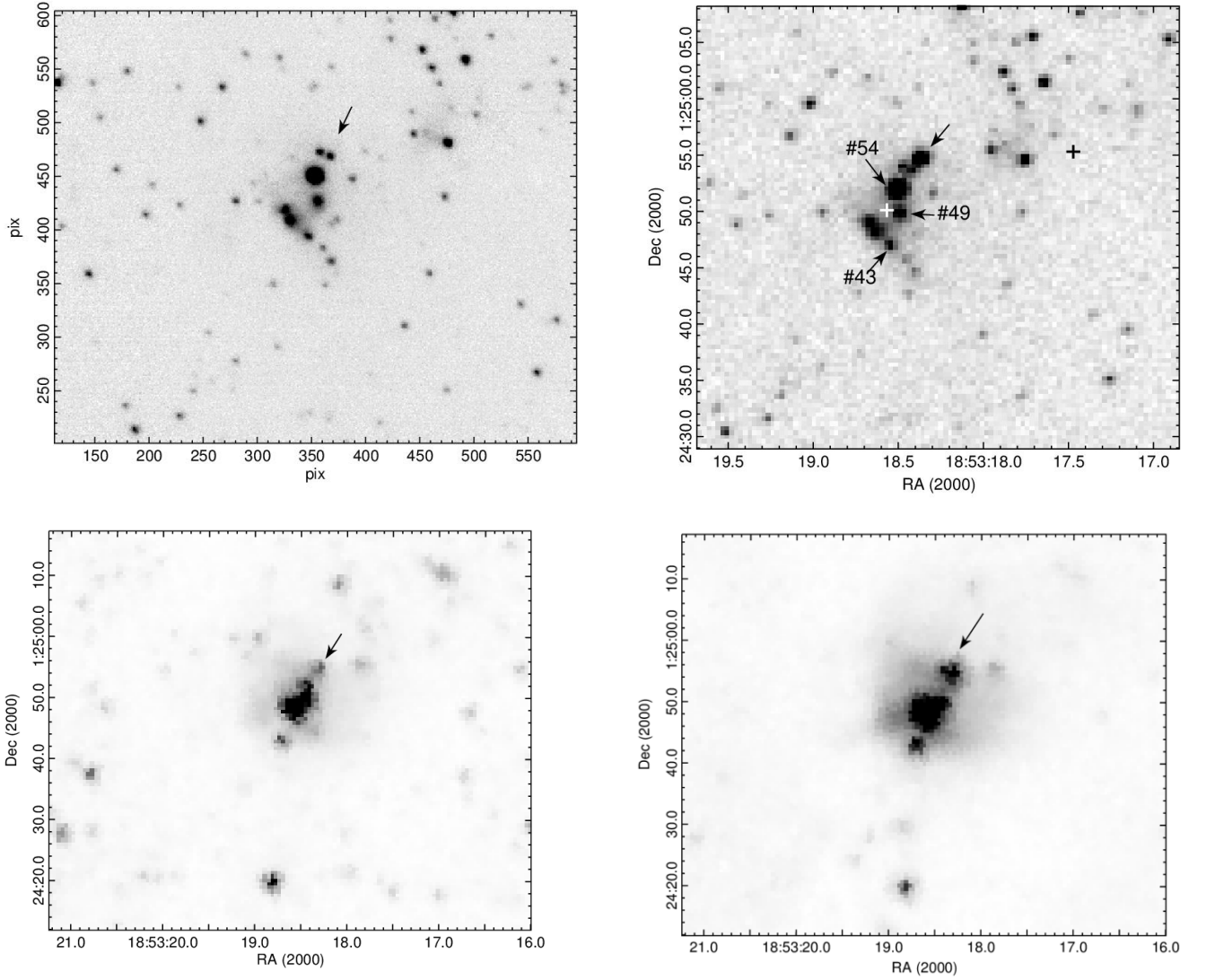

**Fig. 1.** The images of UKIDSS-J185318.36+012454.5 in different wavelengths: Ks band from Varricatt et al (2010) (left top), Ks band from DR6 UKIDSS GPS (right top), [3.6] μm band from GLIMPSE I survey (left bottom) and [5.8] μm band from GLIMPSE I survey (right bottom). The position of IRAS 18507+0121 source from IRAS PSC and IRAS PSC/FSC Combined Catalogue (Abrahamyan et al., 2015) marked by black and white crosses respectively. On DR6 UKIDSS GPS Ks image are marked three stellar sources (#43, 49 and 54, Shepherd et al., 2004), for which in Table 2 the photometric data are presented (a more detailed description is given in the text).

**Table 2.** H and Ks magnitudes of nearby stellar objects

| Object | Varricatt et al. (2010) Ks | DR6 UKIDSS GPS H | DR6 UKIDSS GPS Ks |
|---|---|---|---|
| (#54) UKIDSS-J185318.49+012451.8 | 12.55 ± 0.03 | 13.67 ± 0.01 | 12.55 ± 0.01 |
| (#49) UKIDSS-J185318.48+012449.6 | 14.72 ± 0.04 | - | 14.29 ± 0.01 |
| (#43) UKIDSS-J185318.46+012446.8 | 15.21 ± 0.04 | 17.83 ± 0.07 | 14.83 ± 0.01 |

This conclusion is also confirmed by mid-infrared photometric data presented in Table 3 (see also Fig.1). The first measurement was obtained in 2004, before the outburst detection in the K band. After, the photometric data cover a later period from 2012 to 2014 and indicate the brightness increasing at ~ 1 mag. in [3.6] μm and ~ 0.5 mag. in [4.5] μm bands. We do not consider the MIPSGAL 24 μm data because the located in the immediate vicinity G034.4035+00.23 H II region is saturated, as the majority of RMS sources, that does not allow to do photometric measurements. It should be noted, that all photometric data, which cover a period from 2006 to 2014, show that within the error bar, the brightness of object remained practically at the same level, undergoing only the insignificant fluctuations.

**Table 3.** MID IR photometric data

| Source | Date of obs. | [3.6] [mag] | [4.5] [mag] | [5.8] [mag] | [8.0] [mag] |
|---|---|---|---|---|---|
| GLIMPSE I | 2004 | 11.55 ± 0.35 | 9.62 ± 0.18 | 7.8 ± 0.01 | 6.65 ± 0.09 |
| DEEP GLIMPSE | 2012 | 10.56 ± 0.11 | 9.10 ± 0.04 | − | − |
| AOKEY 4218976 | 2012-06-09 | 10.35 ± 0.30 | 9.04 ± 0.20 | − | − |
| AOKEY 10291200 | 2012-11-16 | 10.54 ± 0.30 | 9.10 ± 0.20 | − | − |
| AOKEY 42728960 | 2014-12-13 | 10.24 ± 0.30 | 9.41 ± 0.20 | − | − |

Unfortunately, the spatial resolution of longer wavelengths images do not allow doing exact measurements of UKIDSS - J185318.36 + 012454.5's fluxes in different wavelengths and epochs. For example, on Fig. 2 the overplots of image in Ks band (UKIDSS GPS) and isophots of sources from MSX (21 μm), WISE 4 (22 μm), SCUBA (850 μm) and ATLASGAL (870 μm) surveys are presented. These sources virtually cover the central part of MM2 clump (Rathborne et al., 2005), which includes a number of young stellar objects overlie the UC H II region produced bright diffuse emission in mid and far-infrared bands (Shepherd et al., 2004, Shepherd et al., 2007, Lumsden et al., 2013). Therefore, it is not possible to determine the exact values of fluxes in different wavelengths separately for the object being studied. The effect of UKIDSS - J185318.36 +012454.5 outburst on the total flux in the longer wavelengths can be insignificant. On the Fig. 3, the broadband continual spectral energy distributions (SEDs) of data obtained before 2003 and after 2006 year are presented. The photometric data we obtained from RMS archive, as well as AKARI and Hi-GAL surveys. For the earlier epoch, before outburst, we used MSX (1996), IRAS (1983), SCUBA (2000) and SIMBA/MAMBO (2001-2002) survey's data, which cover the wavelength range from 8 μm to 1.2 mm. Although the coordinates of IRAS 18507+0121 source according to the IPAC data have the significant offset (~18") from associated with G034.4035+00.23 H II region MSX source, its fluxes also were used. We were guided by the corrected coordinates of IRAS source from IRAS PSC/FSC Combined Catalogue (Abrahamyan et al., 2015), the offset of which from MSX source is only ~ 1.5". For the later period, after the outburst, WISE 12 μm and 22 μm (2010), AKARI (2007), MIPS 70 μm (2010), ATLASGAL (2007), BGPS (2007) and Hershel PACS (2011) were used. They cover practically the same wavelength range from 9 μm to 1.1 mm. Between the integral fluxes obtained in different epochs there is an insignificant difference. In total the flux of later period increased by ~10%, that only slightly exceeds or comparable the error bars of mentioned above photometric data and it cannot be regarded as a clear evidence of the outburst.

The effect of UKIDSS - J185318.36 + 012454.5 outburst on the total flux from H II region is more noticeable in comparing the data obtained in different epochs from relatively equivalent wavelength ranges and bean sizes or equivalent radiuses, such as MSX E and WISE 4 fluxes. The flux of MSX E band (21 μm, a bean size is 18.3") is 13.6 ± 0.8 Jy. The flux in WISE 4 band (22 μm, a bean size is 12") is 17.6 ± 0.2 Jy. The offset between the coordinates is ~1". Thus, the flux on 21 - 22 μm increased by ~4 Jy during the period 1996 - 2010. In this regard, we would like to draw attention to yet one fact. According to the RMS survey with MSX source associated G034.4035+00.2282 HII region and two, identified in UKIDSS GPS, YSOs. One of them is UKIDSS-J185318.36 + 012454.5 or G034.4035+00.2282A. By the authors of Lumsden et al. (2013), its fraction in the total luminosity in MSX wavelengths is ~0.2. For WISE 4 band it equals to 3.5 Jy, that is comparable with the difference between the fluxes. This enables to suggest, that the flux increasing is mostly due to the outburst.

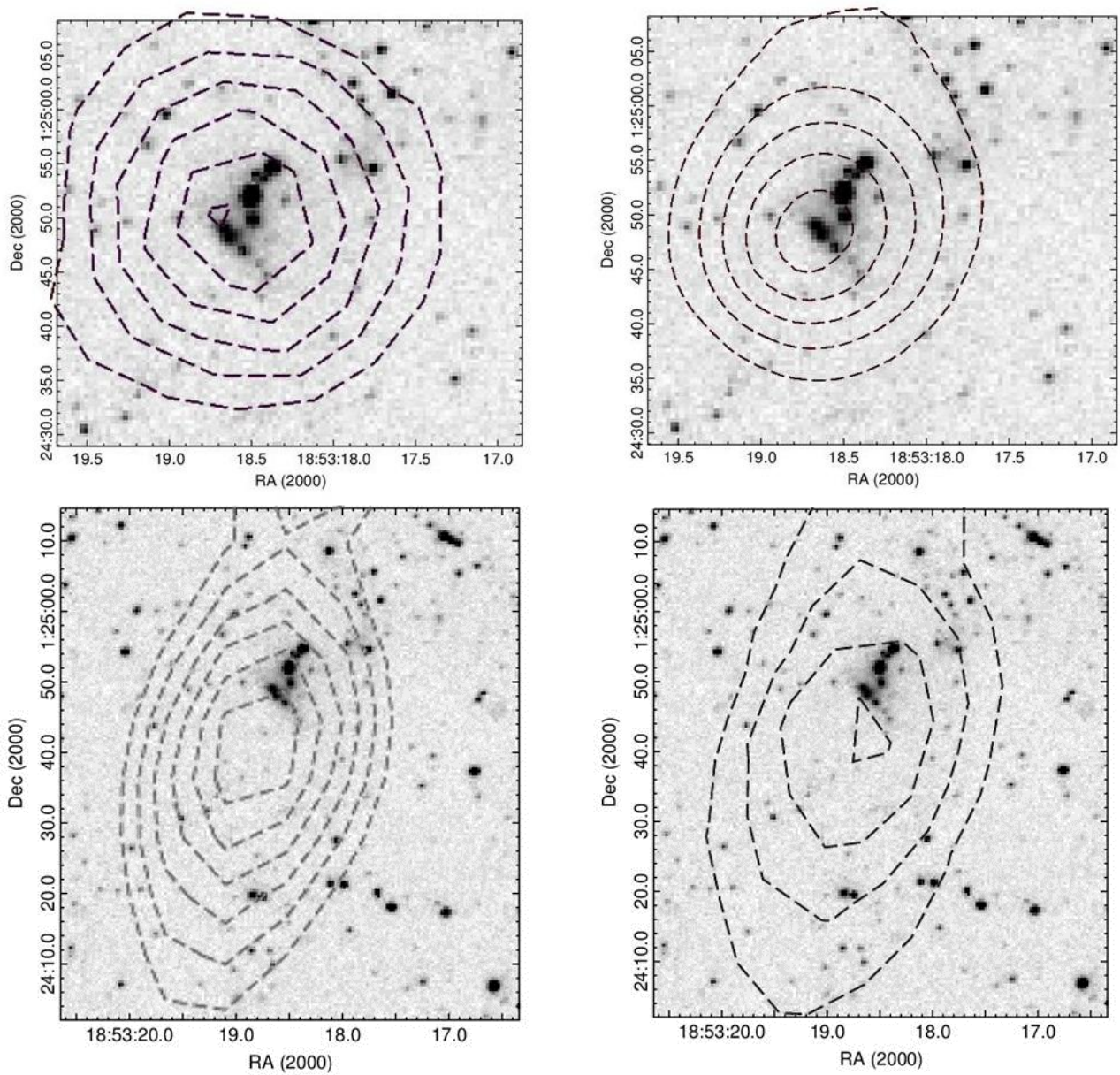

**Fig2.** The overplots of Ks band image with isophots of MSX 21 μm (left top), WISE 22 μm (right top), SCUBA 850 μm (left down) and ATLASGAL 870 μm (right down) sources.

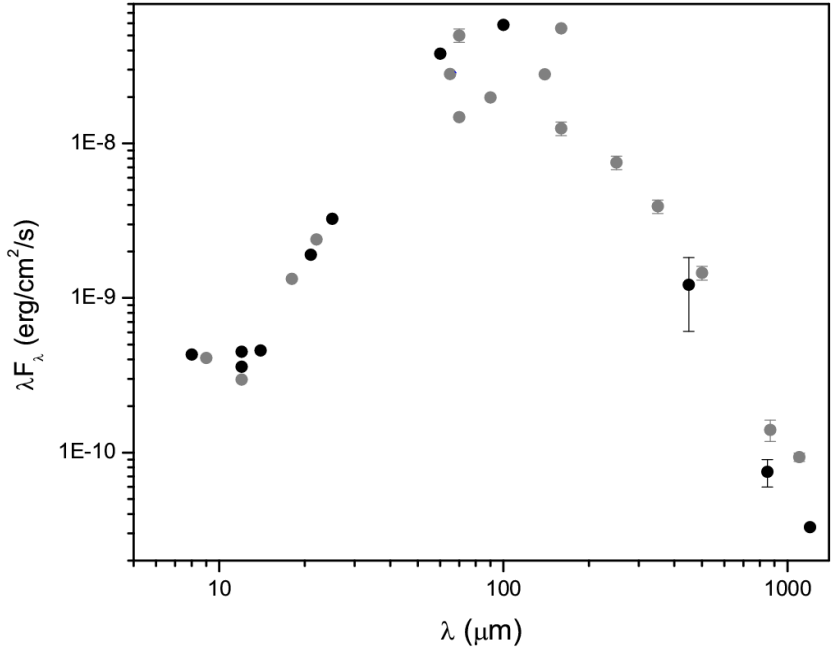

**Fig. 3.** Broadband continuum SED of G034.4035+00.2282 HII region: (black) - the data obtained before 2003 year (MSX, IRAS, SCUBA and SIMBA/MAMBO), (gray) – the data obtained after 2006 (WISE 12 μm and 22 μm, AKARI , Spitzer MIPS, ATLASGAL, BGPS and Hershel PACS.

The increase in brightness can be seen also in a longer wavelengths bands. The flux in SCUBA 850 μm ($R_{ef}$ = 52") is 21.7 ± 4.3 Jy and in ATLASGAL 870 μm ($R_{ef}$ = 40") is 40.5 ± 6.4 Jy. The offset between peak coordinates is ~5.5". Note, that the peak fluxes in both databases are the same and equal to 7.4 Jy/beam. Nevertheless, from our point of view, the rising of the submillimeter flux, integrated with such a large area, can only be regarded as an indirect evidence of the outburst.

*3.2. Spectral properties*

In Cooper et al. (2013), the results of UKIDSS-J185318.36+012454.5 K band spectroscopy are presented. The observation was carried out after an outburst in 2007. The spectral data show the presence of $H_2$ 2.1218 μm and CO emission. The emission of Br series lines, as well as He I 2.0587 μm emission, is below the limit of the detection sensitivity.

The main spectral properties of FUors, which have a higher rise of luminosity and longer duration outbursts, generally display strong CO absorption and mostly lack of emission lines in their infrared spectra. EXors type objects, which have not so strong brightness increasing and shorter outburst duration, show strong emission of HI recombination lines, as well a CO emission during the outburst. The rise of luminosity (ΔK > 5 mag.), outburst duration (near and mid-infrared luminosity shows only slight fluctuations during about 8 years), as well as lack of Br series lines correspond to FUors. On the other hand, CO emission is the one of the main characteristics of EXors. In this regard, we would like to note the following. Among the 19 eruptive variables from the list in Contreas Pena et al. (2016), only in four cases Br series lines are not detected ( VVV v 322, 717, 721 and 815), but in contrast to our object, in the spectra of the first three objects CO observed in absorption. In spectra of the fourth, VVV v815, CO band observed neither emission, nor absorption. Thus, the spectral features of the objects have mixed characteristics of both, FUors and Exors.

*3.3. The evolution stage.*

Fig. 4 shows the two-color diagrams where the positions of UKIDSS-J185318.36+012454.5 according to its near and mid-infrared photometric data obtained in 2004 and after 2006 (see Tables 1 and 3) are marked. The position of the object on the [3.6] - [4.5] versus [5.8] - [8.0] colors diagram allows to classify the objects as Class 0/I YSO (Hartmann et al., 2005). After brightness increasing in the mid-infrared range [3.6] - [4.5] color becomes bluer, that is reflected on K - [3.6] versus [3.6] - [4.5] colors diagram. On this diagram, the object also occupied the position of Class I YSOs

For comparison, on the diagrams the positions of other eruptive variables are marked. The colors of our object are comparable with those of the deeply embedded outburst stellar sources AR6B, OO Ser and GM Cha (Aspin & Reipurth 2003; Hodapp et al. 1996; Persi et al. 2007), which, according their SEDs, classified as Class I eruptive variable YSOs (Gramajo et al., 2014). In addition, the positions of deeply embedded eruptive variables GPS V3, VVVv20, and VVVv815 which are also classified as Class I YSOs with age ~$10^5$ years (Contreras Pena et al., 2014; Contreas Pena et al., 2016) are marked on the diagram. Notice, that VVVv118 (Contreras Pena et al., 2014) with age ~$10^6$ years and nor surrounded by a cool envelope V733 Cep (Reipurth et al., 2007) occupy the position of Class II PMS objects.

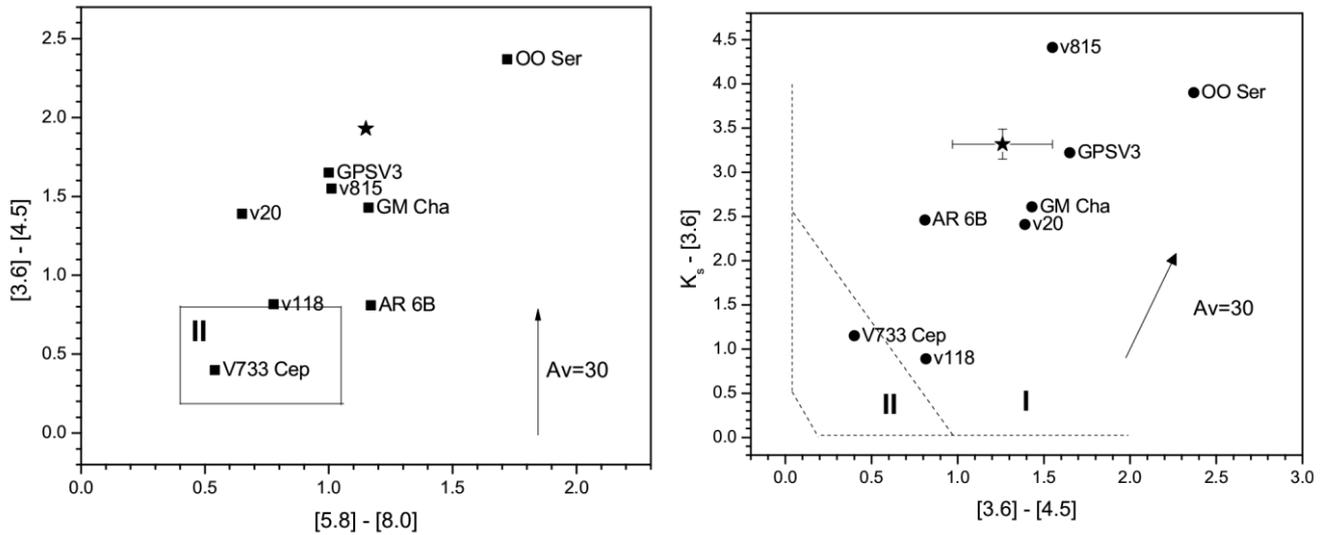

**Fig. 4.** The classification of UKIDSS-J185318.36 + 012454.5 and other eruptive variables using near and mid-infrared measurements. In the left panel, the rectangle marks the domain of Class II sources. The Class I domain is above and to the right (Allen et al., 2004). In the right panel the colors are separated into Class I and II domains by the diagonal dashed line (Gutermuth et al., 2008). Arrows show extinction vectors for $A_V = 30$ (Rieke & Lebofsky, 1985; Flaherty et al., 2007). The position of UKIDSS-J185318.36+012454.5 marked by a star. On the right panel the average magnitudes from Table 1 and 3 are used. The error bars are equal to the standard deviation.

The estimations of an interstellar extinction (Av) value in G034.4035+00.2282 HII region vary from Av=20 mag (Shepherd et al., 2004) to Av=29 mag (according to the COBE/DIRBE and IRAS/ISSA maps in Schlegel et al., 1998). According to the data of observations in $C^{18}O$ line $N_{H2}$ = 2.4 x $10^{22}$ (Xu et al., 2016) and conversation factor between column density $N_{H2} = 9.4 \cdot 10^{20}$ cm$^{-2}$

Av/mag (Rv = 3.1, Bohlin et al., 1978) $A_v$ = 25.5 mag. Apparently, Av ≈ 30 mag. is the upper limit of extinction in this region. Note, that even we take into account the upper limit of interstellar extinction the colors of UKIDSS-J185318.36 + 012454.5 are complied the Class I YSOs.

Combining all the available near- and mid-infrared measurements (H, K, [3.6], [4.5], [5.8], [8.0] and WISE 4), we have constructed the SED of UKIDSS-J185318.36+012454.5, plotted in Figure 5. For the flux on 22 μm, we use the difference between fluxes of MSX E and WISE 4 bands (see text above). The SED of our object is exceptionally red in 1.5 to 5 μm region, with H - [4.5] ≈ 8 mag. At longer wavelengths (from 5.8 to 8.0 μm, before the brightening in mid-infrared) the SED of the object becomes roughly flat. For comparison, the SEDs of other Class I embedded eruptive variables (OO Ser, GM Cha GPS V3 and VVV v815) are plotted in Figure 5. In a mid-infrared region the SEDs' shape of GM Cha, GPS V3 and VVV v815 are similar to UKIDSS-J185318.36+012454.5. But at longer wavelengths, towards to 22 μm, their SEDs decline, that is due to lack of cold dust emission. This characteristic is observed in EXors in outburst phase and explained as arising from a hot inner disk. In long wavelengths range, the SED of UKIDSS-J185318.36+012454.5 is comparable with other embedded eruptive variable OO Ser. The slope of SED $\left(\frac{\Delta log(\lambda F_\lambda)}{\Delta log(\lambda)}\right)$ for 4.5 - 22 μm interval for UKIDSS-J185318.36 + 012454.5 is 1.9 and for OO Ser is 1.7. For both objects, the photometric data of these bands obtained after the outburst. Note, that according to the data presented in Gamajo et al. (2014) the disk mass accretion rate value of OO Ser is the highest of all Class I FUors.

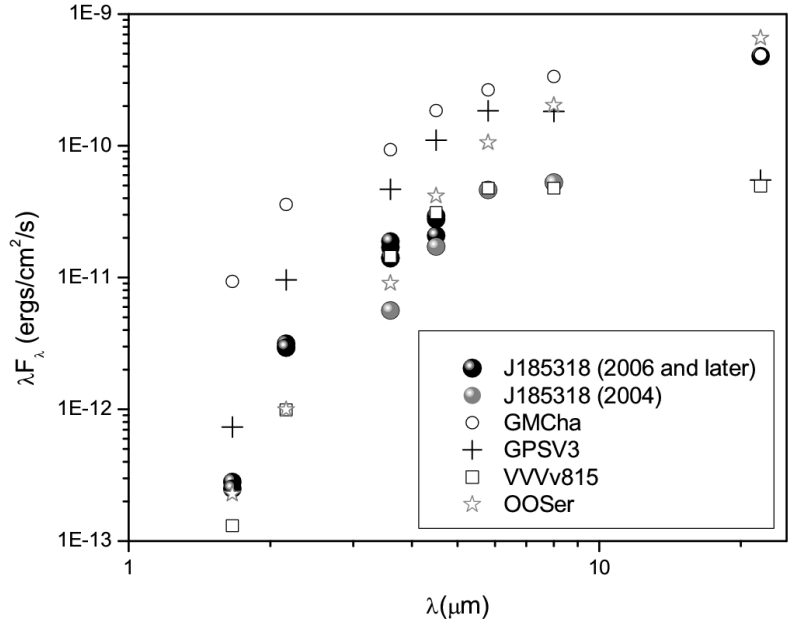

**Fig. 5.** The SEDs of UKIDSS-J185318.36+012454.5, OO Ser, GM Cha, GPS V3 and VVV v815 eruptive variables. For UKIDSS-J185318.36+012454.5 all the photometric measurements from Tables 1 and 3 are marked. For 22 μm the differences of fluxes of MSX E and WISE 4 bands were used, which obtained before and after an outburst (a more detailed description is given in the text).

## 3.4. SED fitting analysis

Thus, according to the near and mid-infrared photometric data the stellar object UKIDSS-J185318.36+012454.5 can be classified as YSO with Class 0/I evolution stage and $\leq 10^5$ years age (Lada & Lada, 2003; Allen et al., 2007). Its evolution stage and location toward the direction to the central part of the MM2 clump of the G034.4035+00.2282 HII region, where it was revealed a number of protostars with ~ $10^5$ years age (Shepheld et al, 2004), suggests, that the object belongs to this star-forming region. The estimations of the distance to this star-forming region are controversial. The trigonometric maser parallax distance determined in Kurayama et al. (2010) is equal to 1.56 kpc. The estimations of a kinematic distance considerably exceed the previous value and vary from 3.7 kpc (Simon et al. 2006) to 3.9 kpc (Molinari et al. 1996). Although a parallax measurement would seem to be more reliable distance determination, the authors of Foster et al. (2012) suggested that the parallax determinations to the same sources are incorrect because of the low declination of this target. All this greatly complicates the classification of the studied variable YSO.

To provide a rough estimation of the basic parameters of the object being studied, we used the SED fitting tool of Robitaille et al. (2007). For the SED fits, as in section 3.3, we used the average near-infrared photometric data from Tables 1, average magnitudes of [3.6] and [4.5] bands obtained after brightening (after 2004, see Table 3) and magnitudes of [5.8] and [8.0] bands obtained in 2004. The SED fitting tool might not be able to give the distinct information for highly variable sources. In addition, the near and mid-infrared photometric measurements are not contemporaneous. To minimize the problem, 10 % uncertainty was assumed for each band. For flux on 22 μm, we use the difference between fluxes of MSX E and WISE 4 bands. Since we have not reliable information about the long wavelength range, we fix the distance ranges from 1.2 to 1.8 kpc and 3.0 to 5.0 kpc for the near and far estimations of distance respectively. As mentioned above, the value of the interstellar extinction (Av) from different sources varies from 20 to 30 magnitudes. We used some wider range from 10 to 40 magnitudes for both cases. Figure 6 and Table 4 show the results obtained after using the SED fitting procedure.

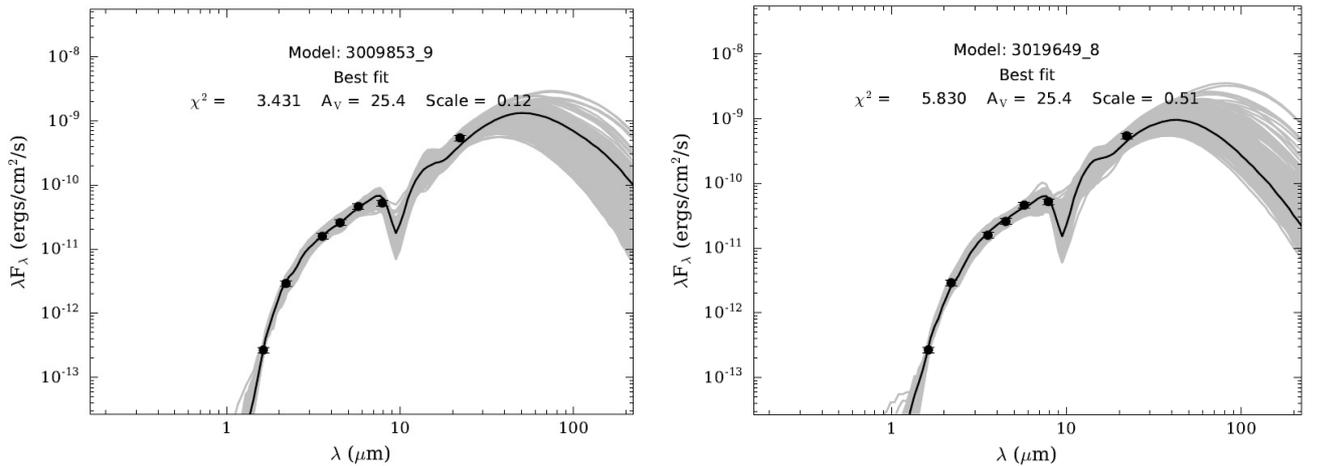

**Fig. 6.** Observed SED and their best fit models for the distance range from 1.2 to 1.8 kpc (left panel) and from 3.0 to 5.0 kpc (right panel). Filled circular symbols represent the data points.

**Table 4.** Parameters derived from Robitaille et al. (2007) models SED fitting

| Parameter | D = 1.2 - 1.8 kpc | | D = 3.0 - 5.0 kpc | |
|---|---|---|---|---|
| | $\chi^2_{best}$=3.4 | $\chi^2$-$\chi^2_{best}$ < 3N | $\chi^2_{best}$=5.83 | $\chi^2$-$\chi^2_{best}$ < 3N |
| Av (mag) | 25.4 | 29.8 ± 4.0 | 25.4 | 29.0 ± 4.0 |
| D (kpc) | 1.32 | 1.47 ± 0.19 | 3.21 | 3.94 ± 0.55 |
| Age ($10^5$ yr) | 0.64 | 0.74 ± 0.33 | 0.24 | 1.55 ± 0.8 |
| Stellar mass ($M_\odot$) | 3.88 | 3.97 ± 0.60 | 7.52 | 7.44 ± 1.02 |
| T(K) | 4430 | 4700 ± 670 | 4530 | 8790 ± 2287 |
| Total luminosity ($L_\odot$) | 111 | 158 ± 36 | 547 | 1170 ± 304 |
| Disk accretion rate ($10^{-5}$ $M_\odot$ yr$^{-1}$) | 3e-4 | 0.15 ± 0.12 | 1.3 | 1.0 ± 1.34 |
| Envelope infall rate ($10^{-5}$ $M_\odot$ yr$^{-1}$) | 7.46 | 7.99 ± 3.90 | 8.84 | 16.7 ± 13.9 |

To identify the representative values of different physical parameters the tool retrieved the best fit model and all models for which the differences between their $\chi^2$ values and the best $\chi^2$ were smaller than 3N, where N is the number of the used data points (as suggested in Robitaille et al. (2007)). This approach is taken because the sampling of the models grids is too sparse to effectively determine the minima of the $\chi^2$ surface and consequently obtain the confidence intervals. Table 4 shows the values for different parameters corresponding to models with $\chi^2_{best}$, as well as the weighted averages and the standard deviations of values for all models with $\chi^2$ - $\chi^2_{best}$ < 3N. The number of such models for each target is called the fitting "degeneracy". It is equal to 259 and 246 for near and far distance estimations respectively. Note, it is not always the best fitting model actually represents the data very well (Grave & Kumar, 2009). For far variant, the estimations of the age, temperature and total luminosity of the model with the best $\chi^2$ are significantly less, than the weighted average value for models with $\chi^2$ - $\chi^2_{best}$ < 3N. The situation is similar to the value of a disk accretion rate for near distance. The value of best fitting model is ~$10^3$ times less than the weighted average.

As expected, the mass estimations according to the SED fitting tools for two variants of distance significantly differ. However, in both cases we can classify UKIDSS-J185318.36+012454.5 as intermediate mass YSO with age ≤ $10^5$ years. Even in the first case (near variant) the object exceeds by mass all Class I FUors from the list in (Gramajo et al., 2014). The ratio between the mass, total luminosity, and age of our object, as well as the disk and accretion rates, within the error bars, correlates well with the several embedded eruptive variable YSOs presented in Contreas Pena et al. (2016). There are objects VVV v20 and v815 for nearer variant and VVV v 405, v473 and v800 for more distant. Note, that comparatively the same relationship between the envelope and disk accretion rates also determined for L1551 IRS 5 and GM Cha infrared eruptive variables (Gramajo et al, 2014).

The value of interstellar extinction (Av) is correlated well with the previous estimations.

*4. Discussion*

The analysis of archival data allows to reveal the new variable stellar object UKIDSS-J185318.36+012454.5. The object, according to the near and mid-infrared colors and SED, is YSO with Class 0/I evolution stage. It is located in the massive star-forming region associated with G034.4+00.23 UC H II region and IRAS 18507+0121 source. On the near-infrared images obtained in 1999, 2000 and 2003 years, the object was indistinguishable. Then, since 2006 the brightness of objects was drastically increased. The amplitude of the brightness increasing during 2003-2006 was as least $K_s$ = 5.0 mag. At this level of brightness, with insignificant fluctuations, UKIDSS-J185318.36+012454.5 remained at least until 2011. Note, that even after outburst, it is invisible in J band. The earliest mid-infrared images of the object were obtained in 2004. The subsequent observations during the period of 2012 - 2014 showed that the brightness of the object increased on ~ 1 mag. in [3.6] band. The brightness variability in [4.5] band significantly lower and does not exceed 0.5 mag. The variability of brightness is detected and in longer wavelengths. The flux of associated with G034.4+00.23 UC H II region WISE 4 source (2010) increased compared with MSX E source (1996) on ~4 Jy.

Several physical mechanisms, such as rotation, cool or hot spots (Scholz et al, 2009) accretion driven wind and outflow (Bans & Königl, 2012) can explain the near-infrared variability observed in YSOs. However, these mechanisms often produce short-term variability with amplitude, that is not expected to exceed 1 mag. in K band. The same physical processes produce the variability in mid-infrared bands (Vijh et al., 2009). But, as a rule, this kind of variability does not exceed 0.6 mag. Therefore, the infrared variations larger than 1 mag. are usually associated with eruptive variability (Scholz et al., 2013). Changes in the extinction along the line of sight can produce larger changes in the magnitudes and on longer time scales (Bouvier et al., 2013). But the brightness variability of nearby stars is not significant and does not exceed ~0.5 mag. It can be seen, as an argument in favor of the fact that such a significant brightness increasing of UKIDSS-J185318.36+012454.5 did not occur due to the extinction variability. Hence, we can consider UKIDSS-J185318.36+012454.5 as an eruptive variable. The increase of a brightness, taking into account the photometric data in both near and mid-infrared ranges, is likely to happen between 2004 and 2006. Note, that the authors of Scholz et al. (2013) concluded, that the objects may be candidates of the eruptive variables, than the brightness at 3.6 and 4.5 μm bands is increased by at least 1 mag. Really, the brightness increasing in [3.6] band is ~ 1 mag., but in [4.5] band is lower, only ~0.5 mag. One can assume, that the Spitzer IRAC photometric data of 2004 obtained during the brightness increasing.

The outburst duration with only slight fluctuations in near and mid-infrared bands allows to classify UKIDSS-J185318.36+012454.5 as an eruptive variable of FUor type. But this classification is not entirely consistent with spectral properties. The spectral observations carried out after the outburst in 2007, shows only the presence of $H_2$ 2.1218 μm and CO emission (Cooper et al. 2013). Note, that one of the basic characteristics of FUor type objects is a CO absorption band (Hartmann & Kenyon 1996). On the other hand, in the spectrum of our object the emission of Br series lines are below the limit of detector's sensitivity. The absence of HI recombination lines is typical for FUor type objects. As a rule, in the spectra of the EXos type objects the emission of HI recombination lines is observed (Herbig 2008). In this regard, we can only offer the following explanation. According to the data presented in (Shepheld et al, 2007) the G034.4+00.23 MM2 clump, where UKIDSS-J185318.36+012454.5 is located, is a source of the massive outflows observed in millimeter and mid-infrared wavelengths. It is not excluded that CO absorption of the stellar object is masked by very strong outflow emission. A similar situation is observed in the optical range in the spectra of FUor type object V 2494 Cyg, where H$\alpha$ absorption is masked by very strong and broad jet emission. However, the summation of the photometric and spectral data does not allow to classify UKIDSS-J185318.36+012454.5 as FUor or EXor. We can consider it as an eruptive variable with mixed characteristics, namely, MNor.

Contradictory estimations of the distance of the star-forming region to which, probably, the object belongs, complicate the determination of its mass and luminosity. However, for both distance estimations (the trigonometric maser parallax distance is 1.56 kpc and the kinematic distance is 3.7 - 3.9 kpc) we can classify UKIDSS-J185318.36+012454.5 as intermediate mass YSO with age $\leq 10^5$ years according to the SED fitting tool. The ratio between the mass, total luminosity, and age of the object, as well as the disk and accretion rates, within the error bars, correlates well with the several embedded eruptive variable YSOs presented in (Contreas Pena et al., 2016).

## 5. Conclusion

On the basis of the infrared data and images, we have revealed the new eruptive variable UKIDSS-J185318.36+012454.5. According to the near and mid-infrared colors and SED of the object, we can classify it as YSO with Class 0/I evolution stage. It is located in the vicinity of IRAS 18507+0121 source and probably belongs to the massive star-forming region associated with GAL 034.4+00.23 HII region. The outburst occurred in the period of 2004-2006. The amplitude of brightness is at least $K_s$ = 5.0 mag. At this level of brightness, the object remains at least until 2014. The spectral observations carried out after the outburst shows the presence of $H_2$ 2.1218 μm and CO emission. The emission of Br series lines is below the limit of detector's sensitivity. The summation of the photometric and spectral data does not allow classifying UKIDSS-J185318.36+012454.5 as FUor or EXor. We can consider it as an eruptive variable with mixed characteristics or MNor type object. According to the data obtained by SED fitting tool, UKIDSS-J185318.36 + 012454.5 is an intermediate mass YSO with age $\leq 10^5$ years.

*Acknowledgements.* We are very grateful to the anonymous referee for his/her helpful comments and suggestions. This research has made use the data obtained at UKIRT which is supported by NASA and operated under an agreement among the University of Hawaii, the University of Arizona, and Lockheed Martin Advanced Technology Center; operations are enabled through the cooperation of the East Asian Observatory. We gratefully acknowledge the use of data from the NASA/ IPAC Infrared Science Archive, which is operated by the Jet Propulsion Laboratory, California Institute of Technology, under contract with the National Aeronautics and Space Administration. We thank the colleagues in the GLIMPSE I and DEEP GLIMPSE Spitzer Legacy Surveys. This paper is also based on observations taken as part of program 083.C-0846(A) which is operated by the European Southern Observatory. We also gratefully acknowledge the authors of RMS survey (Lumsden et al., 2013), as well as papers Cooper et al. (2013) and Varricatt et al. (2010).